# Distributed-Something: scripts to leverage AWS storage and computing for distributed workflows at scale.

Erin Weisbart[1] and Beth A. Cimini[1,2]

1. Broad Institute of MIT and Harvard, Cambridge MA, USA; Department: Imaging Platform
2. Corresponding author

On-demand computational infrastructure, such as that provided by Amazon Web Services (AWS), is ideal for at-scale parallelizable workflows (especially workflows for which demand is not constant but comes in occasional "spikes"), as neither computing power nor data storage are limited by local availability and costs are limited to actual resource usage. However, cloud infrastructure configuration is time-consuming and confusing, and cloud-native services that automatically monitor and scale resources can increase the workflow price. Distributed-Something (DS) is a collection of easy to use Python scripts that leverage the power of the former, while minimizing the problems of the latter.

DS makes it possible for a user with moderate computational comfort to design a way to deploy a new tool or program to the cloud, and for a user with relatively low computational comfort to then deploy this tool at will. It simplifies the process of using AWS by scripting the majority of the setup, triggering, and monitoring of jobs, requiring only minimal human readable config files to be edited before the run, and simple, single-line commands to trigger each step. This increases the ability of novice computationalists to be able to execute workflows and dramatically lowers the barrier to creating a new workflow. Unlike most existing tools for running hosted containerized analysis, it does not require learning new workflow languages (either for new-tool-addition or end-user-deployment) and minimizes the understanding requirement of the AWS components that are used (see Supplementary Table 1).

We originally sought to simplify large scale scientific image analysis using our CellProfiler software, creating Distributed-CellProfiler[1]. Recognizing the utility of the framework, we herein provide Distributed-Something as a fully customizable template for the distribution of any Dockerized[2] workflow. We show its extensibility with two example implementations of DS in the open source bioimage ecosystem (though DS is in no way limited to bioimage analysis).

ImageJ is the most widely used open-source software for bioimage analysis[3]; Fiji is an open-source distribution of ImageJ that comes bundled with libraries and plugins that extend ImageJ's functionality[4]. Fiji scripts can be run at scale using Distributed-Fiji, allowing the user to take advantage of its plugin ecosystem and its ability to run user-written scripts in many coding languages. As with all DS implementations, the computational environment can be tailored to each task (e.g. many small machines used to individually process thousands of files or a large machine to perform a single task on many files (such as stitching)).

To increase shareability of especially large bioimage data, the Open Microscopy Environment[5] team is creating next generation file formats, including .ome.zarr[6], to make bioimaging data more findable, accessible, interoperable and reusable (FAIR)[7]. We created Distributed-OmeZarrCreator to simplify the conversion of large bioimage datasets to .ome.zarr's and thus encourage the adoption of this format and simplify sharing of bioimaging data via resources such as the Image Data Resource (IDR)[8].

DS coordinates 5 separate AWS resources. Data is stored on AWS in its Simple Storage Service (S3) and "Spot Fleets" of Elastic Compute Cloud (EC2) instances (or virtual computers) access that data, run the "Something" on that data, and upload the end product back to S3. ECS (Elastic Container Services) places your customized Docker containers on the EC2 machines while Simple Queue Service (SQS) tracks the list of jobs, and Cloudwatch provides logs and metrics on the services you are using, allowing for configuration optimization and troubleshooting. One can easily customize DS code to download or upload data from/to cloud and/or on-premises storage outside the AWS account used for processing.

DS shines in the simplicity of end-user execution: only two human-readable files must be edited to configure individual DS runs: the Config file and the Job file. The Config file contains information about naming, the number and size of machines to use, and the maximum price you are willing to pay for the machines, minimizing computational costs. The Job file lists all of the individual tasks to run in parallel by setting both metadata shared between tasks and the metadata to parse individual tasks. An additional Fleet file contains information about AWS-account-specific information but does not need to be edited after initial creation.

Three single-line python commands initiate all of the AWS architecture creation and coordination and an optional fourth command provides additional monitoring and automated clean-up of resources (detailed in Figure 1).

Implementing your own version of DS can be done in a matter of hours by someone with moderate Python abilities. Creating a Distributed- version of a software that itself takes input scripts (e.g. Distributed-Fiji) makes workflow customization possibilities near limitless with no extra overhead. Over 1000 containers are already registered on BioContainers[9] and, conceptually, any could be put in the DS framework.

We believe DS will enable the scientific community to quickly, easily, and cost-effectively scale their parallelizable workflows using AWS. As this is an open-source tool, we look forward to contributions and implementations from within and outside the bioimage analysis community.

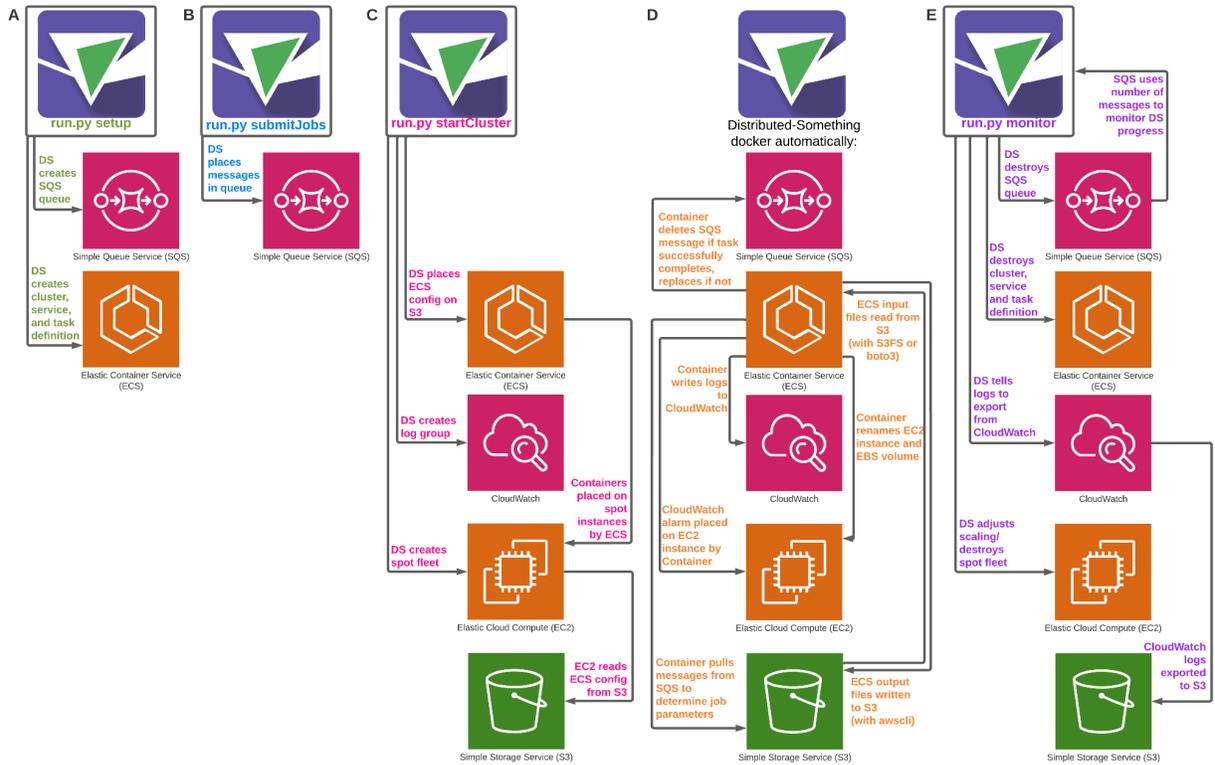

## Figure 1

Distributed-Something uses four single-line commands to coordinate five separate AWS services for the parallel processing of jobs by any Dockerized software. Three execution commands prepare various aspects of AWS infrastructure. `setup` (A) prepares SQS and ECS. `submitJobs` (B) sends jobs to SQS. `startCluster` (C) initiates and coordinates the spot fleet request. After these commands, the Distributed-Something Docker automatically completes setup (D) and jobs run. A fourth optional command, `monitor` (E), assists in downscaling and cleaning up resources as they are no longer required. See online documentation at https://distributedscience.github.io/Distributed-Something for a deeper discussion of each step.

Supplementary Table 1

| Software | End user input format | Cost | Strengths | Weaknesses | Code Availability | Complete workflow examples available | Ease of use for developers | Ease of use for end users |
|---|---|---|---|---|---|---|---|---|
| Distributed-Something | - Human readable metadata files | Compute plus monitoring ($0.0001/hour/machine) | - New tools can be added with only Python in ~1 hr<br>- Automatic job AND container-level introspection, tagged by job parameters<br>- Automatic machine monitoring | - CLI only<br>- Only perfectly parallel single-stage tasks<br>- Infrastructure teardown requires a script, must be run by the user or set on an automatic schedule | Open-source, single codebase | Few | Moderate | Easy |
| Terra | - Workflow Description Language | Compute | - Uses data stored in Google Cloud, Azure, or AWS<br>- Web interface<br>- Run batch jobs or interactive notebooks | - Workflow language required for end-users | Open-source, across many codebases | Many | Moderate | Moderate |
| Galaxy | - Drag-and-drop nodes | None to end users (cost | - Very easy for end users<br>- Automated | - Scaling dependent on bandwidth of your Galaxy host | Open-source, across many codebases | Many | Challenging | Easiest |

| | | borne by hosters) | logging and historical introspection<br>- Run batch jobs or interactive notebooks | - Creating a wrapper can be challenging for new developers due to custom XML | | | | |
|---|---|---|---|---|---|---|---|---|
| AWS Batch | - Individual jobs can be submitted in console<br>- Scripting needed to add jobs at scale | Compute only, monitoring optional | - Automatically optimizes compute settings<br>- Set up infrastructure via console or CLI | - No built-in container-level introspection<br>- More expensive cluster-level introspection due to need for custom metrics<br>- Documentation encourages Fargate over EC2 which is ~10% more expensive. | Proprietary | Few | Moderate | Moderate |
| AWS Parallel Cluster | - Scripting | Compute only, monitoring optional | - Automatically optimizes compute settings if using AWS Batch<br>- Easy to move to for slurm users if using slurm | - If using slurm, more difficult for end-users to monitor job progress (requires SSH into the head node)<br>- If using AWS Batch, see above | Open source, single codebase | Few | Moderate | Moderate |
| Amazon Genomics CLI | - Workflow Description Language<br>- Nextflow | Compute plus "overhead" ($4/day) | - Easy connection to AWS Registry of Open Data (RODA) resources | - Workflow language required for end-users<br>- Less control of the infrastructure used for jobs | Open source, single codebase | Few | Moderate | Moderate |

# Code Availability

Distributed-Something is available at https://github.com/DistributedScience/Distributed-Something
Distributed-CellProfiler is available at https://github.com/DistributedScience/Distributed-CellProfiler
Distributed-Fiji is available at https://github.com/DistributedScience/Distributed-Fiji
Distributed-OmeZarrCreator is available at https://github.com/DistributedScience/Distributed-OmeZarrCreator

# Acknowledgements

We thank Juan Caicedo and Shantanu Singh for creating the original Distributed-CellProfiler, Callum Tromans-Coia, Josh Moore, and Sébastien Besson for their assistance with DOZC, and other members of the Cimini and Carpenter-Singh labs for their feedback on this project and manuscript. This study was supported by Calico Life Sciences LLC, NIH P41 GM135019, and grant number 2020-225720 from the Chan Zuckerberg Initiative DAF. The funders had no role in study design, data collection and analysis, decision to publish or preparation of the manuscript.

# Author information

## Authors and Affiliations

1. Broad Institute of MIT and Harvard, Cambridge MA, USA; Department: Imaging Platform
   Beth A. Cimini, Erin Weisbart

## Contributions

B.A.C. conceived the project, wrote DS and DF, assisted in writing DCP, and revised the manuscript.
E.W. wrote the manuscript, wrote DOZC, and assisted with DS and DF.

## Corresponding author

Correspondence to Beth Cimini

# Ethics declarations

## Competing interests

The authors declare no competing interests.